# HOT CARRIER SOALR CELLS : IN THE MAKING ?

**A. Le Bris[1]*, L. Lombez[1], JF. Guillemoles[1], R. Esteban[2], M. Laroche[3], JJ. Greffet[3], G. Boissier[4], P. Christol[4], S. Collin[5], JL. Pelouard[5], P. Aschehoug[6], F. Pellé[6]**
[1]Institute of R&D on Photovoltaic Energy (UMR 7174, EDF-CNRS-ENSCP), 6 Quai Watier BP 49, F78401 Chatou
*corresponding author: arthur.le_bris@centraliens.net
[2]CSIC-UPV/EHU and DIPC, 20018 Donostia-San Sebastian, Spain
[3]LCFIO, Institut d'Optique, 2 avenue Augustin Fresnel, F91127 Palaiseau
[4]IES, Université de Montpellier 2, F34095 Montpellier
[5]LPN, route de Nozay, F91460 Marcoussis
[6]LCMCP, Chimie Paristech, 11 rue Pierre et Marie Curie, F75231 Paris

ABSTRACT: Hot carrier solar cells allow potential efficiency close to the thermodynamical limit in ideal conditions. However, the feasability of such devices has not been clearly stated so far and only ideal cells were considered in previous studies. Here we develop a model with realistic energy selective contacts, carrier thermalization and absorptivity. The requirements in term of contact selectivity is investigated, showing that semi-selective contacts are not incompatible with high efficiencies. Candidates for absorbing material were synthesized and the required thermalization properties are obtained. Specific structures were designed to enhance absorption of concentrated sunlight in a small material thickness, allowing high carrier density in the absorber.

## 1 INTRODUCTION

The hot carrier solar cell (HCSC) is an attractive concept allowing a potential efficiency close to the thermodynamical limit [1], with a relatively simple structure (see figure 1). In such a cell, the carriers are kept in a thermal disequilibrium with the lattice, thanks to a reduced electron-phonon interaction [2]. They are collected through energy selective contacts, allowing only carriers having one specific energy to be collected, to prevent a heat flow upon carrier extraction towards a cold population [3].

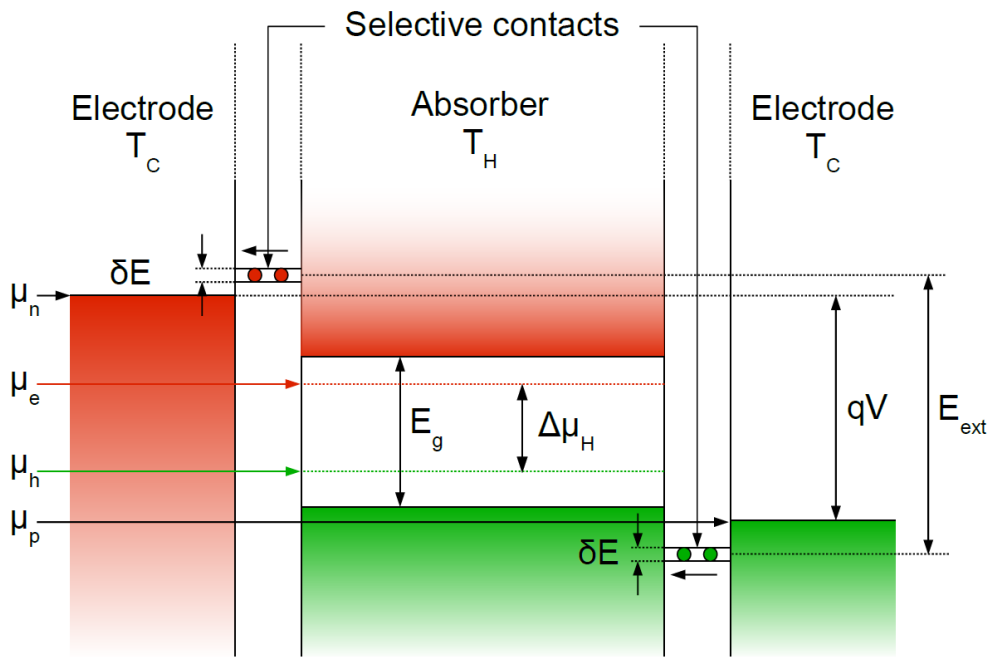

**Figure 1**: Schematic of a hot carrier solar cell. Electron-hole pairs are photogenerated in the absorber where they are kept "hot" (at temperature $T_H$). They are extracted through selective contacts having a small transmission range δE towards a cold distribution in the electrodes (at temperatrue $T_C$).

Models of ideal HCSC were proposed [1,3] showing a potential efficiency up to 86 % under a fully concentrated (~46000 times) black body spectrum. However, the feasibility of such a device has not been clearly stated so far. In particular, achieving the required carrier thermalization rate in the absorber may be challenging. Also, theoretical and experimental study on selective contacts have shown the difficulty to obtain good selectivity and high conductivity [4].

Here, simulations are used to determine the required carrier thermalization properties and contact selectivity for a hot carrier effect to occur. A material and device structure is proposed to achieve the targeted carrier cooling rate. Also, a design is proposed to obtain high absorption in a small material thickness to approach the high carrier density that is necessary.

## 2 THERMAL LOSSES THROUGH CONTACTS

Simulations are used to determine the effect of the contact non ideality on the cell efficiency.

### 2.1 Model

The model used is based on a detailed balance model [1,5], where the charge (or energy) current extracted is the difference between the absorbed photon (or energy) flux and the emitted photon (or energy) flux. Absorption and emission are described by a generalized Planck law [6]. In previous models, the contacts were assumed ideal with a discrete transmission probability. In that case, the energy collected per extracted carrier is that of the carrier, and is equal to the energy difference between the hole contact and the electron contact $E_{ext}$. Minimal losses upon carrier extraction are obtained in that case.

Here, we considered a non zero transmission range δE. In that case, if the energy $E_{ext}$ + 2δE is extracted from the absorber, only $E_{ext}$ is usable to produce work, the rest being lost as heat. A thermalization process occurs at the contacts that increases when increasing the transmission range.

The charge and energy current delivered by the absorber is then defined as follows [7,8]:

$$J_{el} \propto \int_{E_{\min}}^{E_{\max}} \tau(E)\left[f_H(E)-f_C(E)\right]dE \qquad (1)$$

$$P_{el} \propto \int_{E_{\min}}^{E_{\max}} E\tau(E)\left[f_H(E)-f_C(E)\right]dE \qquad (2)$$

where $J_{el}$ and $P_{el}$ are the electric and energy current density delivered by the cell, $E_{min}$ and $E_{max}$ are the lower and upper bound of the contact transmission range, τ(E) is the transmission probability of a carrier having energy E, and $f_C(E)$ and $f_H(E)$ are the carrier distribution function of the cold and hot population respectively. In this paper, τ(E) will be taken unity in the contact transmission range, zero outside.

By solving the charge and energy balance equations using equations (1) and (2) [8], the current is determined as a function of the voltage, and the efficiency is obtained for a given δE.

## 2.2 Results of simulations

Simulations are made considering a 1 eV band gap absorber under fully concentrated 6000K black body spectrum. Two cases are considered : an ideal case where thermalization is negligible, and a more practical case with thermalization losses in the absorber. In the latter, a thermalization factor of Q=1 W/K/cm² is considered, which is ten times lower than that of a multi quantum wells GaAs sample [9,10] (see section 3.1).

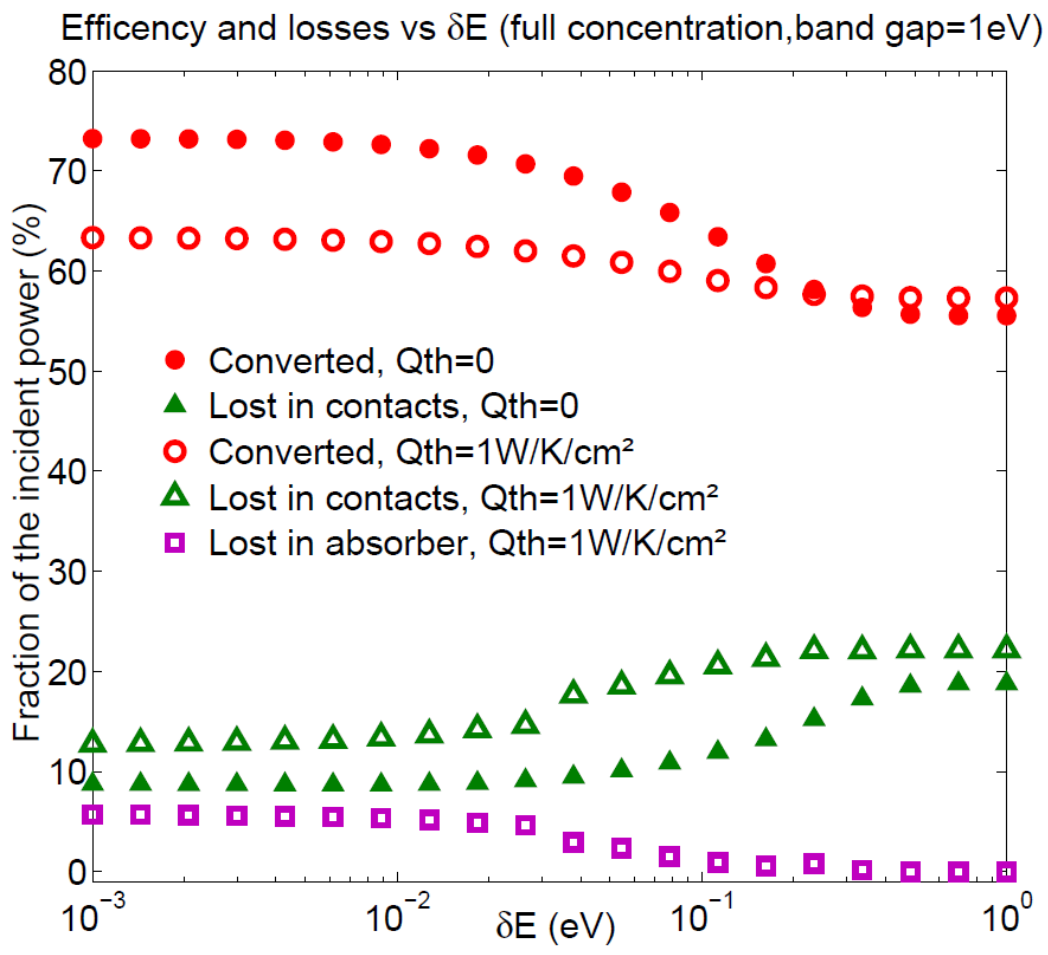

**Figure 2**: Efficiency (circles) and losses (in the absorber: squares, and in the contacts: triangles) of a HCSC as a function of the contact transmission range. Two thermalization factor are considered: negligible thermalization (filled symbols), and Q=1W/K/cm², corresponding to a thermalization rate ten times lower than MQW GaAs.

Figure 2 presents the efficiency and losses of a HCSC as a function of the contact transmission range and for the two thermalization factors that are considered here. For $\delta E<10^{-2}$ eV, corresponding to δE small compared to $kT_H$, the contact can be considered ideal, with an efficiency that is constant and close to the optimal value. Above $10^{-2}$ eV, the losses in contacts increase, causing a drop of efficiency. However, for $\delta E>>kT_H \sim 0.1$ eV, the efficiency becomes independant of the transmission range. In that case, the contacts are only semi-selective, allowing carrier extraction above a threshold energy $E_{ext}$. One can see that in the limit of broad transmission range, the efficiency is still above the Shockley-Queisser limit. This show that semi-selective contacts are compatible with efficiency enhancement which may allow a much simpler design.

# 3 REDUCTION OF CARRIER THERMALIZATION

## 3.1 Results of simulations

Carrier thermalization in the absorber is taken into account in the simulation by adding thermal losses in the energy balance.

$$P_{el}=P_{abs}-P_{em}-P_{Th} \quad (3)$$

where $P_{el}$ is the electric power, $P_{abs}$ and $P_{em}$ are the absorbed and radiatively emitted power, and $P_{Th}$ is the power lost by thermalization. The thermal losses term is expressed is considered here as a function of the carrier temperature [8,9]:

$$P_{Th}=Q\left(T_H-T_C\right)\exp\left(\frac{-E_P}{k_B T_H}\right) \quad (4)$$

$T_H$ and $T_C$ being the temperature of the carriers and of the lattice respectively, $E_P$ is the energy of the zone center LO phonon, and Q is a thermalization factor and a material constant.

One can see on figure 2 that a HCSC having a thermalization factor of 1W/K/cm² (ten times lower than a GaAs multi quantum well sample) would give a potential efficiency around 60%. Such a value is our target for future devices.

## 3.2 Experimental study

We investigated antimony based compounds for their good absorption properties, and for their small band gap that is suitable for HCSC application. Also, they are expected to have a low thermalization rate.

It has been demonstrated [10] that nanostructured materials exhibit a reduced thermalization rate (see figure 3) compared to bulk. This comes from a reduction of the phonon density of states that limits the thermalization pathways. Also, phonon confinement may prevent heat flow outside of the absorber, and could help keeping the carrier population excited. For that reason, we studied quantum wells samples.

Two types of samples have been studied: strained samples (V812) made of 5x10 nm thick InGaSb wells and 20 nm thick GaSb barriers, and lattice matched samples (V725) made of 5x10 nm thick InGaAsSb wells and 20 nm thick AlGaAsSb barriers. The samples were made at IES.

We determined the cooling rate of these samples using continuous wave photoluminescence. By looking at the carrier temperature (from the high energy side of the PL peak) as a function of the laser power, it is possible to determine the material thermalization factor [11].

Results obtained at low temperature (50 K) are encouraging, with estimated values of the thermalization factor around 2 W/K/cm², close to the targeted value. However, a confirmation of this result at ambient temperature is necessary.

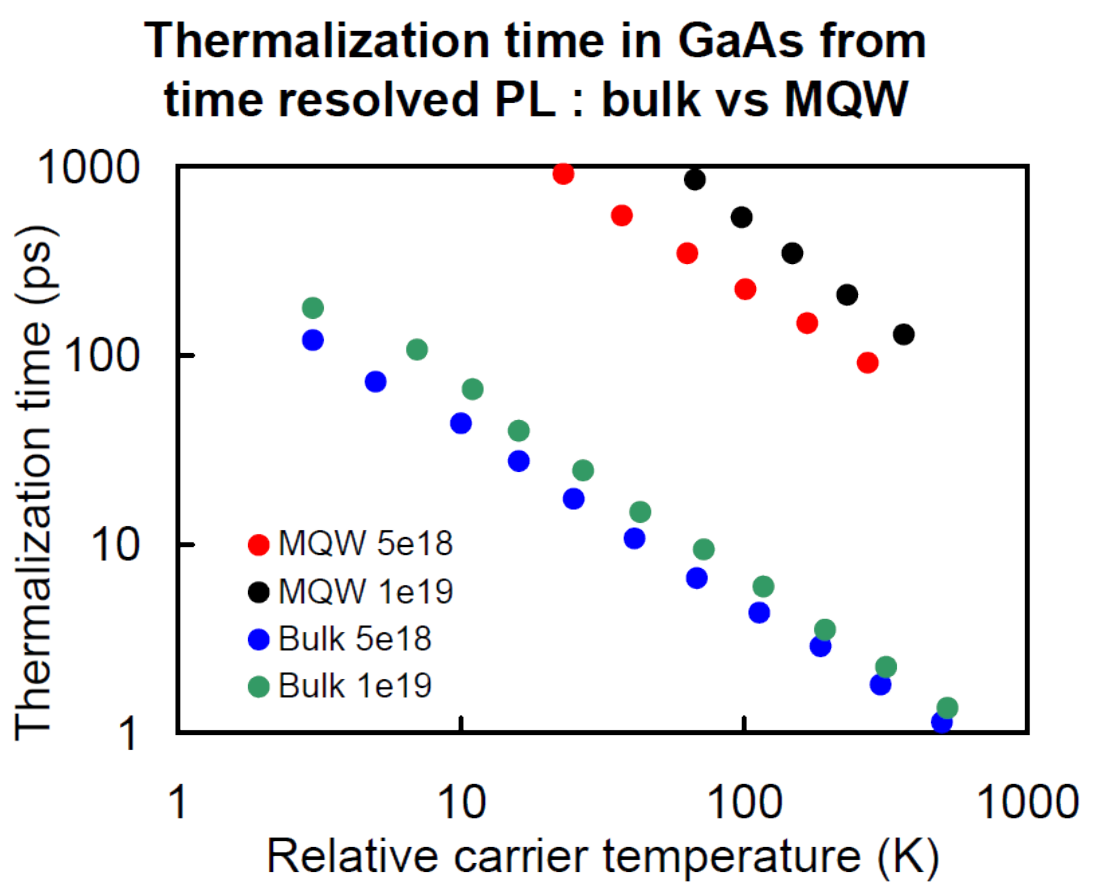

**Figure 3**: thermalization vs carrier temperature of bulk (green and blue dots) and quantum wells (red and black dots) GaAs. Two carrier densities are considered: 5e18cm$^{-3}$ and 1e19cm$^{-3}$

3.3 Light absorption enhancement

We can see on figure 3 that the thermalization time can be lengthened with increased carrier densities. With higher carrier density per volume unit, it becomes easier to saturate the thermalization pathways which is related to the number of available phonon modes.

This means that achieving a high concentration of the solar energy allows to reduce the thermalization properties requirement. It is therefore useful to be able to efficiently absorb the incident light under concentration in a reduced thickness of material.

For that purpose, resonant dielectric structure are designed to enhance the absorption of the solar spectrum over a wide spectral and angular range in a 50nm thick GaAs absorber. The proposed structure in figure 4 is made of a GaSb layer sandwiched between a dielectric 1D saw tooth grating and a silver mirror [12].

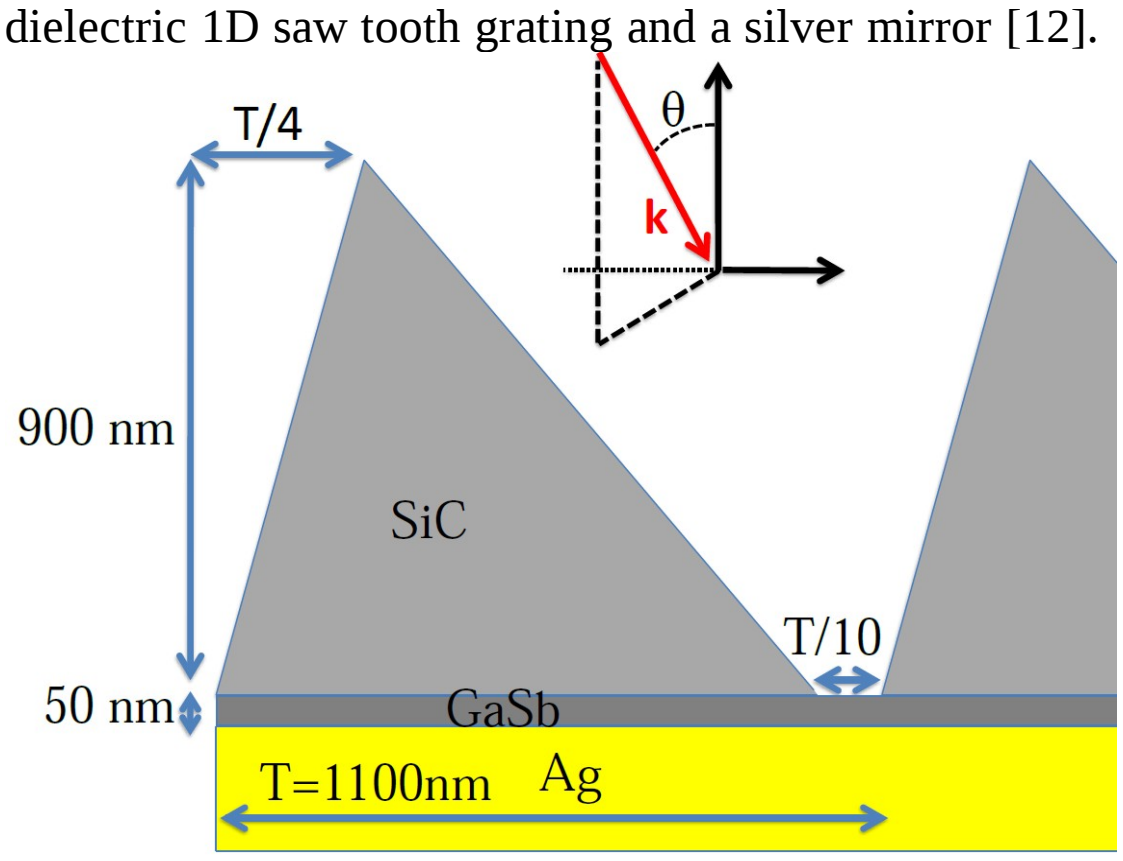

**Figure 4:** Resonant dielectric structure for enhanced light absorption over a wide spectral and angular range. The absorbing 50 nm thick GaSb layer is sandwiched between a SiC grating and a silver back reflector.

Simulations predict 66% absorption of the total incident photon flux below 1800 nm under ~10000 suns ($\theta$ < 45°) in the GaSb layer, which represents a 76% increase compared to the same layer without the saw tooth structures.

## 4 CONCLUSION

Different technological locks in the HCSC device have been addressed. Resonant structures at the front are proposed and are expected to allow efficient absorption in a thin absorbing layer. The material and structure of the studied samples give promising thermalization properties that make it a good candidate as an absorber. Finally, simulations show that semi-selective contacts enable efficiency up to 60%, higher than the Shockley-Queisser limit.

We thank all our partners of the THRI-PV project (IES, ILV, LCFIO, LCMCP, LGEP, LPN) and GCEP and ANR programs for financial support.